\documentclass[11pt,english]{article}
\usepackage{ae,aecompl}
\usepackage{helvet}
\usepackage[T1]{fontenc}
\usepackage[authoryear]{natbib}
\usepackage{amssymb,amsmath,amsthm}
\usepackage{graphicx}
\usepackage[pdfborder={0 0 0}]{hyperref}
\usepackage{chngpage}
\usepackage{pdflscape}
\usepackage{epstopdf}
\usepackage{fullpage}
\usepackage{setspace}
\usepackage{booktabs}
\usepackage[small,compact]{titlesec}
\usepackage{rotate}
\usepackage{subcaption}

\usepackage{hyperref}

\usepackage{multirow}
\usepackage{array}
\usepackage{graphicx}

\newcommand\MyBox[2]{
	\fbox{\lower0.75cm
		\vbox to 1.2cm{\vfill
			\hbox to 1.2cm{\hfil\parbox{1.2cm}{#1\\#2}\hfil}
			\vfil}%
	}%
}

\theoremstyle{plain} 
\theoremstyle{plain} 
\theoremstyle{plain} 

\bibpunct[: ]{(}{)}{;}{a}{,}{,}

\begin{document} 
	
	\title{Pursuing Open-Source Development of Predictive Algorithms: The Case of Criminal Sentencing Algorithms}
	
	\author{Philip D. Waggoner\footnote{Corresponding author}\\University of Chicago\\ \texttt{pdwaggoner@uchicago.edu} \and Alec Macmillen\\University of Chicago\\ \texttt{amacmillen@uchicago.edu}}
	
	\date{ }
	
	\maketitle

	\begin{abstract}
		\noindent Currently, there is uncertainty surrounding the merits of open-source versus proprietary algorithm development. Though justification in favor of each exists, we argue that open-source algorithm development should be the standard in highly consequential contexts that affect people's lives for reasons of transparency and collaboration, which contribute to greater predictive accuracy and enjoy the additional advantage of cost-effectiveness. To make this case, we focus on criminal sentencing algorithms, as criminal sentencing is highly consequential, and impacts society and individual people. Further, the popularity of this topic has surged in the wake of recent studies uncovering racial bias in proprietary sentencing algorithms among other issues of over-fitting and model complexity. We suggest these issues are exacerbated by the proprietary and expensive nature of virtually all widely used criminal sentencing algorithms. Upon replicating a major algorithm using real criminal profiles, we fit three penalized regressions and demonstrate an increase in predictive power of these open-source and relatively computationally inexpensive options. The result is a data-driven suggestion that if judges who are making sentencing decisions want to craft appropriate sentences based on a high degree of accuracy and at low costs, then they should be pursuing open-source options.
	\end{abstract}

\vspace{0.3in}

\textbf{No conflict of interest}

\vspace{0.3in}

	
	\clearpage
	
	\doublespacing
	
Algorithmic recidivism prediction is an increasingly popular topic in the criminal justice system as well as in the computational community. These statistical models can aid human actors like judges in making crucial decisions about bail and sentencing, significantly reducing the financial and human capital burdens of caseload processing. Currently, there is uncertainty surrounding the merits of open-source versus proprietary development of these algorithms. For example, proprietary development could be considered preferable because market-based drivers like competition promote effective development, and confidential criminal justice data might be better protected in proprietary settings. Yet, on the other hand, it could be the case that open-source collaboration reduces costs and encourages the transparency necessary to ensure that predictive risk assessment does not propagate bias against marginalized identity groups, in addition to wider technical engagement.

There exist several proprietary predictive risk assessment products currently in use, including COMPAS, LSI-R, and PSA. Perhaps the best-known is COMPAS (Correctional Offender Management Profiling for Alternative Sanctions). COMPAS is owned and maintained by Equivant, a private software company, and provides risk scores for a given offender’s pretrial release risk, general recidivism risk, and violent recidivism risk. COMPAS has been deployed in states across America, including California, Florida, New York, and Wisconsin. The LSI-R (Level of Service Inventory – Revised) is similar tool that quantifies an offender’s risk of re-offending and identifies the offender’s particular rehabilitation needs. LSI-R has been implemented in American states including Iowa, Kansas, and Rhode Island, and also abroad, in Australia. Finally, PSA (Public Safety Assessment) provides scores that a given individual will recidivate if released pending trial, and the risk that individual will fail to appear for future court hearings. PSA was developed by Arnold Ventures, a limited-liability corporation that funds philanthropic and political organizations and projects.

Noted in the previous citations, there are reasonable arguments on both sides of the open-source versus proprietary development issue. However, we suggest that in sensitive, real-world applications that significantly affect human lives, open-source development should be the standard for reasons of transparency, collaboration, and cost-effectiveness. The case of criminal sentencing algorithms is particularly well-suited to demonstrate this argument. In particular, decisions like bail-setting and sentencing are highly consequential for defendants. Integrating statistically efficient computational modeling into criminal justice decision-making has legal, ethical and philosophical implications, and to appropriately parse these thorny issues, we argue that open-source development should be the industry standard. 

Algorithmic recidivism prediction is a particularly timely topic because recent studies have uncovered racial bias in proprietary algorithms like COMPAS \citep{angwin2016machine, dressel2017accuracy, dressel2018accuracy}. Furthermore, issues of over-fitting and model complexity likely plague these proprietary models’ estimates, if indeed a classifer with only two features can produce substantively identical results as a classifer with 137  \citep{dressel2018accuracy}. The opaque processes by which the proprietary options are developed makes these algorithms difficult to challenge or amend, because they are effectively black boxes where the relationship between inputs and outputs is unclear. 

To address these issues and in service of our argument, we replicate a major proprietary algorithm using real criminal profiles and subsequently fit three penalized regressions to predict and classify recidivism post-sentencing. We then demonstrate an increase in the predictive accuracy of these open source and relatively computationally inexpensive options. Our analysis demonstrates that relatively simple open-source options are preferable to proprietary risk assessment algorithms. An open-source approach, then, should result in predictions with accuracy just as good as or better than proprietary models at far lower cost, making open-source the superior development approach.

\section{Pursuing Open Source Predictive Algorithm Development}

Open-source development of risk assessment software is superior to a proprietary, closed-source approach for several reasons. Chief among these are the benefits of collaboration, reduction of costs, and transparent, ethical development. Commercial recidivism prediction tools like COMPAS are developed in private and protected as trade secrets. They can also be costly, which precludes widespread adoption, especially in under-resourced jurisdictions. Finally, and most problematically, because such commercial tools are essentially black boxes, it is difficult to evaluate and correct for any potential biases they may propagate. An open-source approach to developing, deploying and applying predictive algorithms would address each of these concerns.
 
One of the many advantages of open-source development is that it encourages collaboration from many perspectives. When developers and researchers from multiple disciplines work together, the project benefits from their diverse skillsets and experience. So too is the case in algorithm development. As such, the task of developing and applying existing predictive sentencing algorithms in the context of criminal recidivism is likely best tackled by interdisciplinary teams of researchers from a range of social scientific and computational backgrounds. Experts in fields ranging from criminology and sociology to statistics and machine learning can pool their knowledge bases to ensure well-rounded solutions flow directly from theoretical as well as technical expertise. And such development in an open-source way ensures that many voices will retain contribution access, thereby strengthening development over time.

Another value of open-source algorithm development involves costs. Proprietary tools like COMPAS incur expenses that may be prohibitive for often smaller, less well-funded jurisdictions that might otherwise benefit from the assistance of automated recidivism prediction. On the other hand, researchers can use free, open-source statistical software to produce models that perform just as well as if not better than expensive commercial alternatives. These tools can then be distributed to under-resourced court systems, especially in rural areas, for a fraction of the cost of existing solutions and with relatively minimal investments needed for training. Open-source software can also reduce the burden of human capital costs on the criminal justice system. Human assessments of recidivism likelihood require the decision maker to acquire years of experience, and such predictions remain vulnerable to bias and error. Of course, algorithmic approaches are only as good as the data and statistical methods that constitute them, and they are not a complete substitute for human expertise. Rather than entirely replace human judgment, we suggest that algorithmic predictive tools can augment the process by which judges make decisions (e.g., setting sentences and bail). Similar to the use of autopilot by airline pilots, judges in this open-source world can employ a hybrid decision-making assisted but not supplanted by risk assessment software. Such a human-in-the-loop approach would enable judges to make better-informed decisions more efficiently while still maintaining control over the process and minimizing financial costs. 

Open-source collaboration also encourages transparent and ethical development. Recidivism prediction is a particularly sensitive application of machine learning because it can have serious impacts on human lives. If judges are basing their sentencing and bail decisions on algorithmic predictions, it is critical that these predictions are not only interpretable and accurate, but also developed above the surface to encourage greater scrutiny. An open-source approach to developing prediction tools ensures that mistakes and systematic biases are identified and corrected quickly. Otherwise, risk assessment tools can have adverse effects on historically marginalized groups by making prediction errors that are correlated with traits like race and gender \citep{dressel2018accuracy}. It is also important that other researchers, judges, and defendants alike are able to understand and, when appropriate, challenge the methods and data that underlie such significant predictions, with the ultimate decision still remaining with the judge. Without transparency in development, these algorithms are effectively black boxes, which not only undermines public trust but also precludes opportunities for performance improvement. There could also be implications for defendants’ legal rights, as demonstrated in Loomis v. Wisconsin, in which a criminal defendant sentenced by COMPAS alleged that its closed-source nature violated his due process rights because he was unable to challenge its scientific validity and accuracy.\footnote{\url{https://www.nytimes.com/2016/06/23/us/backlash-in-wisconsin-against-using-data-to-foretell-defendants-futures.html}; \url{https://www.scotusblog.com/wp-content/uploads/2017/02/16-6387-cert-petition.pdf}} Adopting an open-source development framework would address these concerns. 

While the benefits of open-source development for risk assessment algorithms are numerous, it is important to note possible drawbacks as well. For example, potential privacy violations pose a challenge. The data used to train risk assessment models often contain personal information included in criminal records. Thus, fully open-source development may be considered tantamount to publicly releasing criminal records for thousands of citizens. Related, open-source development provides no guaranteed continuity of support. With proprietary software, payment ensures that users will have access to resources that assist with training and usage, and that the software will actually be maintained over time. Developers are not typically compensated for their work on open-source projects, making it possible that these tools could eventually lose support due to attrition for financial reasons and become outdated. This scenario could result in a patchwork of partially-complete tools, each lacking individuals or organizations dedicated to keeping them development-ready.

Although these critiques are valid, we believe there are simple steps that could be taken to address them. First, regarding privacy, the training data used to build the risk assessment models could be anonymized so that the individuals are not identifiable. To ensure that the models are continuously updated with new data, courts across the country could be encouraged to periodically contribute anonymized data to an open repository. Learners could be trained on this public data but deployed in confidential contexts to protect defendants’ privacy in real time. Next, regarding dedicated support, it is important for collaborating researchers to proactively manage and delegate responsibilities to ensure open-source risk assessment tools are kept current. A unified framework for pursuing financial resources (whether through research grants or private funding) could be pursued. 

There are undoubtedly challenges to an open-source approach to algorithmic recidivism prediction, but with proper planning we suggest they can certainly be overcome. In light of the recent surge in research on risk assessment algorithms \citep{eckhouse2019layers, green2019disparate, holsinger2018rejoinder}, social implications \citep{dressel2018accuracy, olhede2018growing}, and combining human decision making with machine advice in public spaces \citep{kennedy2018trust, waggonerbig}, the conditions may be right to attract support to mitigate against the drawbacks of open-source development, especially given the mounting costs (both financial and otherwise) of proprietary development. 

\section{Empirical Strategy}

To support our argument that open-source algorithm development is an optimal path forward for designing and deploying predictive algorithms, especially in a consequential public space, for the remainder of the paper we are interested in empirically demonstrating this point. We do so by fitting three widely used predictive models using open-source statistical software and demonstrate the resulting increase in accuracy. These models are three types of penalized regressions and are popular in modern data science and machine learning applications: ridge regression, LASSO regression, and elastic net regression. Each of these methods has been refined through extensive contribution from many researchers and each can be inexpensively implemented using open-source statistical software. An additional benefit of these models is that they have desirable statistical properties that produce more precise and accurate predictions of outcomes. We explore these properties in further depth below. These penalized regression models lend themselves to frequent, transparent deployment and improvement, making them ideal for producing fast, efficient and statistically strong solutions in important public applications like recidivism prediction. Such an approach reinforces our central argument that open source development invites a host of contributions from the best minds acctively working in the field.

In this application, we use real criminal profile data from Broward County, FL at the heart of the recent controversy surrounding proprietary sentencing algorithms \citep{angwin2016machine}. These data are valuable, not only because they are real criminal profiles useful for testing different algorithms in a real-world setting, but they allow us to speak directly to this controversy and thus contribute to the conversation.

In service of this goal, we proceed in three steps. First, we replicate the predictions from COMPAS algorithm following previous approaches in the literature \citep{dressel2017accuracy}. With COMPAS as a baseline, we then fit our open-source models, which are statistically well-suited for such a task. Finally, we demonstrate that our relatively straightforward, open-source, and relatively computationally inexpensive approach yields more accurate predictions of the likelihood of recidivism for these real people, compared to the COMPAS approach. Such a process strengthens our core argument that better, transparent solutions can and do exist for vexing and consequential public problems. 

Prior to fitting the models, we first offer a brief introduction to these models to motivate our empirical approach that follows. Penalized regressions are becoming increasingly prominent in the modern machine learning and data science landscapes. These models were have a rich history with roots in the ridge regression developed several years ago \citep{hoerl1970ridge}, but with recent updates including the LASSO regression \citep{tibshirani1996regression}, and, most recently, the elastic net regression \citep{zou2005regularization}.\footnote{Importantly, though these are the three penalized regressions most often cited and used (hence our starting place here), there are numerous exentions of the penalized logic, but in different data contexts such as the Bayesian LASSO \citep{park2008bayesian}, the kernel ridge regression \citep{orsenigo2012kernel}, and penalized regression for heterogeneous feature selection \citep{wu2010heterogeneous}. } Their contemporary resurgence is due in large part to their ability to efficiently classify outcomes by penalizing large, redundant regression coefficients, which also allows for sorting between competing theoretical claims. To the first point, large regression coefficients are undesirable because they contribute to model instability and sub-optimal predictions. Any statistical model asserts a theoretical relationship by the mere inclusion of predictors on the right-hand side of the equation. The combination of feature predictors $\{X_1\dots X_p\}$ are thought to explain some unique amount of variance in the response $Y$. In the social sciences, a researcher includes specific features in a model specification to express an internally consistent theory and testable expectations in the form of a hypothesis. The latter point concerning penalized regression's utility is relevant where multiple theoretical claims compete with one another, implying an over-specified, saturated model is fit to explain the outcome $Y$. In such cases, penalized regression offers a statistically efficient avenue for sorting between these distinct claims, because it drops features or tunes coefficient estimates to reflect the relative importance of their unique impacts on the response $Y$. 

Penalized regressions do this with a simple modification to the loss function, where either the $\ell_{1}$-norm or the $\ell_{2}$-norm penality is added to the function, such that overly large coefficients contributing to model instability are either minimized or dropped completely from the specification.\footnote{The $\ell_{1}$ norm of a coefficient vector is defined as, $\normalsize \|\beta\|_{1} = \sum_{j=1}^{p}|\beta_{j}|$, where the $ \ell_{2}$ norm of a coefficient vector is defined as,
$\|\beta\|_{2} = \sqrt{\sum_{j=1}^{p}\beta_{j}^{2}}$.} As this is not a technical paper and as these models have been thoroughly scrutinized, deployed, and extended in a host of data science contexts, we present only the base loss function to highlight the inclusion of the penalty and to demonstrate how this affects coefficient estimates for each regression type.

Notably, in a penalized regression context, the goal is to minimize the training error, where we are interested in obtaining penalized coefficient estimates, $\hat{\beta}_{\lambda}^{X} $, that minimize information loss. The first of these three is the ridge regression, where the penalty parameter, $\lambda$, is applied to the some vector of squared coefficients,

\begin{equation}
\hat{\beta}_{\lambda}^{ridge} = argmin \{\sum_{i=1}^{n} \bigg\lgroup y_{i} - \beta_{0} - \sum_{j=1}^{p}\beta_{j}X_{ij} \bigg\rgroup ^{2} + \lambda\sum_{j=1}^{p}\beta_{j}^{2}\}.
\label{eq:ridge}
\end{equation}

In Equation \ref{eq:ridge}, note the first term is the base loss function and the added $\ell_{2}$-norm penalty, tuned by $\lambda$, is $\lambda\sum_{j=1}^{p}\beta_{j}^{2}$. Such a penalty shrinks the coefficient estimates to be very small, but stops short of dropping them from the specification. As such, the second approach to penalizing over-specified models is the LASSO regression picks up at this point and employs the $\ell_{1}$-norm penalty (also tuned by $\lambda$),

\begin{equation}
\hat{\beta}_{\lambda}^{LASSO}  = argmin \{\sum_{i=1}^{n} \bigg\lgroup y_{i} - \beta_{0} -\sum_{j=1}^{p}\beta_{j}X_{ij} \bigg\rgroup ^{2} + \lambda\sum_{i=1}^{p}|\beta_{j}|\}.
\label{eq:lasso}
\end{equation}

Note the subtle but powerful change in penalty in Equation \ref{eq:lasso}. Here, the absolute value of the coefficient estimates are now penalized, $\lambda\sum_{i=1}^{p}|\beta_{j}|$. This has the effect of dropping redundant features from the specification, thereby strengthening the overall fit and predictive performance of the model. This is the distinction with the ridge case in Equation \ref{eq:ridge}. 

Finally, we turn to the third approach, which is often called elastic net regression. This approach combines the $ \ell_{1} $ and $ \ell_{2} $ norms by adding the penalities together,

\begin{equation}
\hat{\beta}_{\lambda}^{EN}  = argmin \{\sum_{i=1}^{n}\bigg\lgroup y_i-\beta_0-\sum_{j=1}^{p}\beta_{j}x_{ij}\bigg\rgroup ^2+\lambda_1 \sum_{j=1}^{p}|\beta_{j}|+\lambda_2 \sum_{j=1}^{p}\beta_{j}^{2}\}.
\label{eq:en}
\end{equation}

However, though added together, the elastic net regression is actually concerned with finding the optimal blend of these penalties to shrink some features and drop others. As such, an additional tuning parameter, $\alpha$, is needed to determine the degree to which these parameters should be blended. Here, as with all penalized regressions, optimality is defined as the specification with the smallest loss of information measured as mean squared error. Thus, we update Equation \ref{eq:en} accordingly, 

\begin{equation}
\hat{\beta}_{\lambda}^{EN}  = argmin \{\sum_{i=1}^{n}\bigg\lgroup y_i-\beta_0-\sum_{j=1}^{p}\beta_{j}x_{ij}\bigg\rgroup ^2+\lambda(\frac{1-\alpha}{2}\sum_{j=1}^{m}\hat{\beta}^{2}_{j}+\alpha\sum_{j=1}^{m}|\hat{\beta}_{j}|)\}
\label{eq:en2}
\end{equation}

In sum, the impact of the penalty parameter, $\lambda$, which constrains the estimated coefficients to either be dropped (absolute value constraint) or minimized (squared value constraint) is clear across these approaches. The proper tuning of these models then, through cross validation, allows for maximally predictive models, which is discussed in the following subsection.

\subsection{Cross-Validation for Model Tuning}

The optimal value for $\lambda$ is most commonly selected through a process called $k$-fold cross-validation. In this process, which we employ in this paper, the data are subdivided into $k$ (in our case, 10) smaller, equal-sized data sets. Each subset is further divided into a training set usually comprising 80\% of the fold, and a testing set that includes the remaining 20\% of the fold. Each training subset is then used to fit a base model, and each test subset is used to generate predicted outcomes at many levels of $\lambda$. There are multiple methods for selecting the optimal $\lambda$ from this distribution of possibilities, including simply choosing the $\lambda$ that gives the minimum error. In this application, we select the $\lambda$ value that is one standard deviation greater than the minimum $\lambda$ in line with other common approaches \citep{zou2005regularization}. This tuned regularization parameter determines the extent to which coefficients are shrunk (ridge regression), dropped (LASSO regression), or both shrunk and dropped (elastic net regression).

Note that in the elastic net case previously mentioned that there is a blending of the $ \ell_{1}$ and $ \ell_{2}$ penalities. Thus, the elastic net regression requires an additional tuning parameter, $\alpha$, which controls this blending, resulting in some compromise between the ridge and LASSO models. To find the optimal $\alpha$ value, we follow the same 10-fold cross-validation strategy for many values for $\alpha$, across many values of $\lambda$. The search culminates in the selection of the optimal values of $\lambda$, which is the value that falls one standard deviation above the minimum mean squared error, and $\alpha$ that, with $\lambda$, jointly produce the model with sufficiently minimal mean squared error.

The penalized regression framework is particularly well-suited for our case of predicting the likelihood of recidivism as a function of a vector of theoretically relevant covariates and criminal history. The ridge, LASSO and elastic net are widely used in similar applications because they produce optimal classifications while minimizing information loss and tuning error. These methods represent an excellent starting place to demonstrate the value of open-source development of recidivism prediction models, especially in comparison to the less transparent and less accurate proprietary iterations. As such, in this paper we are interested in building a suite of optimal, widely used open source classifiers that maximize predictive performance, while including only necessary features that contribute to the overall fit of the model.

\subsection{Data \& Features}

As previously noted, we use real criminal profile data from Broward County, FL \citep{angwin2016machine}. These data include 7,214 criminal records and a follow-up feature on whether the defendant committed another crime within two years. If they did, this would be considered recidivism. The other important features in the data we use in this analysis include the features included in Dressel's (2017) thesis, which is the basis of the Dressel and Farid (2018) piece, and is also replicated in the appendix. Thus, for consistency and comparative ability, we use the features she used. First, we include the defendants' race, which is a continuous indicator where 1 = ``white'', 2 = ``black'', 3 = ``hispanic'', 4 = ``asian'', 5 = ``native american'', and 6 = ``other.'' Next we included the sex of the defendant, which was binary where 1 = ``female'' and 0 = ``male''. Then we included continuous indicators for defendants' ages, juvenile felony counts, juvenile misdemeanor counts, juvenile ``other'' crime classification counts, and counts of all non-juvenile crimes. The final feature is a binary indicator for whether a crime was a misdemeanor (0) or a felony (1). 

\section{Analysis}

In this section, we present our results in four ways: model-level predictive accuracy rates, model-level classification accuracy, feature-level tuned coefficient plots, and model-level accuracy via receiver operating characteristic (ROC) curves and area under the curve (AUC) calculations.

\subsection{Predictive Accuracy}

In the first stage of presenting our results, we start by presenting mean accuracy prediction rates across 1,000 iterations of all specifications in Table \ref{table:simple.tab1}. Each cell entry is the mean accuracy rate across all iterations of each model, with the exception of the COMPAS replication accuracy rate which is in the first row.\footnote{The COMPAS replication did not require mutliple model iterations as the data included a unique column feature for the COMPAS prediction, allowing for a precise baseline.} 

An important feature in these data allowing us to ``replicate'' the COMPAS finding is the \textit{compas prediction} feature, which is binary with a 1 suggesting the COMPAS algorithm predicted recidivism and a 0 suggesting the COMPAS algorithm predicted no recividism within two years. For replication, we compared the predictions to the observed cases of recidivism and found an accuracy rate of 65.4\%. This acts as the baseline against which we compare the predictive accuracy of our three penalized regressions. The expectation is that if our prediction accuracy rates are higher, then we can have strong evidence that our open-source approach is both statistically efficient, as well as more accurate than the COMPAS algorithm.

Seen across all models below the COMPAS row in Table \ref{table:simple.tab1}, the mean accuracy rates were indeed around 2\% higher for each. Specifically the mean accuracy rates for 1,000 iterations of the LASSO, ridge and elastic net models were 67.2\%, 67.1\%, and 67.4\%, respectively. The most accurate across the many iterations was the elastic net, which is corroborated by the single specification presented and discussed in the third stage in Figure \ref{figure:main}. The consistency across these models is also notable. While each approach penalizes model complexity differently, the results are robust across each specification, meaning this is a reasonable place to begin to a discussion on the value of open-source, statistically efficient alternatives to proprietary sentencing algorithms. Ultimately this is a useful first step in presenting our analysis, suggesting that relatively straightforward, widely accessible solutions are possible. 

	\begin{table}[h!]
	\begin{center}
		\caption{Average Accuracy Rate by Model Type}
		\label{table:simple.tab1}
		\begin{tabular}{ |l c| }
			\hline
			Model & Average accuracy rate \\
			\hline
			\hline
			COMPAS (baseline) & 65.4\% \\
			LASSO & 67.2\% \\
			Ridge & 67.1\% \\
			Elastic net & 67.4\%\\
			\hline
			\multicolumn{2}{r}{\textit{Note: Average accuracy rate from 1,000 iterations of model fit and evaluation.}}
		\end{tabular}
	\end{center}
\end{table}

\subsection{Classification Accuracy}

In the second stage, we present an alternate description of model accuracy with confusion matrices for the classification solutions across all models in Tables \ref{table:confLASSO}, \ref{table:confRIDGE}, and \ref{table:confEN}.\footnote{Note that all cell entries may be just shy of adding up to 100\% due to rounding. But these cell entries reflect the overall and proportional accuracy of classifying recidivists.} Here, we are interested in higher \textit{true} values, whether positive or negative, suggesting our models did a better job classifying true cases, relative to the false classifications. 

First note the relatively stable classification errors across all models. For example, the true positive rates for the LASSO, ridge, and elastic net models are 22.6\%, 23\%, and 23.2\%, respectively. As such, it is clear from these results that there were consistently higher proportions of true classes correctly classified compared to false classes, suggesting the algorithms did well in classifying recidivism, at least relative to incorrectly classifying criminals post-sentencing.

	\begin{table}[h!]
	\begin{center}
		\renewcommand{\arraystretch}{1.1}
		\caption{Confusion Matrix: LASSO}
		\label{table:confLASSO}
		\begin{tabular}{c > {\bfseries}r @{\hspace{0.7em}}c @{\hspace{0.4em}}c @{\hspace{0.7em}}l}
			& \multicolumn{4}{c}{\bfseries Observed outcomes} \\
			\multirow{10}{*}{\rotatebox{90}{\parbox{1.1cm}{\bfseries\centering Predicted\\ outcomes}}} & & \bfseries p & \bfseries n & 
			\\
			& p$'$ & \MyBox{\centering 326\\(22.6\%)}{} & \MyBox{\centering 122\\(8.5\%)}{} & 
			\\[2.4em]
			& n$'$ & \MyBox{\centering 314\\(21.8\%)}{} & \MyBox{\centering 680\\(47.2\%)}{} & 
			\\
		\end{tabular}
	\end{center}
\end{table}

\begin{table}[h!]
	\begin{center}
		\renewcommand{\arraystretch}{1.1}
		\caption{Confusion Matrix: Ridge}
		\label{table:confRIDGE}
		\begin{tabular}{c > {\bfseries}r @{\hspace{0.7em}}c @{\hspace{0.4em}}c @{\hspace{0.7em}}l}
			& \multicolumn{4}{c}{\bfseries Observed outcomes} \\
			\multirow{10}{*}{\rotatebox{90}{\parbox{1.1cm}{\bfseries\centering Predicted\\ outcomes}}} & & \bfseries p & \bfseries n & 
			\\
			& p$'$ & \MyBox{\centering 332\\(23.0\%)}{} & \MyBox{\centering 130\\(9.0\%)}{} & 
			\\[2.4em]
			& n$'$ & \MyBox{\centering 308\\(21.4\%)}{} & \MyBox{\centering 672\\(46.6\%)}{} & 
			\\
		\end{tabular}
	\end{center}
\end{table}

\begin{table}[h!]
	\begin{center}
		\renewcommand{\arraystretch}{1.1}
		\caption{Confusion Matrix: Elastic Net}
		\label{table:confEN}
		\begin{tabular}{c > {\bfseries}r @{\hspace{0.7em}}c @{\hspace{0.4em}}c @{\hspace{0.7em}}l}
			& \multicolumn{4}{c}{\bfseries Observed outcomes} \\
			\multirow{10}{*}{\rotatebox{90}{\parbox{1.1cm}{\bfseries\centering Predicted\\ outcomes}}} & & \bfseries p & \bfseries n & 
			\\
			& p$'$ & \MyBox{\centering 335\\(23.2\%)}{} & \MyBox{\centering 128\\(8.9\%)}{} & 
			\\[2.4em]
			& n$'$ & \MyBox{\centering 305\\(21.2\%)}{} & \MyBox{\centering 674\\(46.7\%)}{} & 
			\\
		\end{tabular}
	\end{center}
\end{table}

\subsection{Exploring the Feature Level: Visualizing Tuned Coefficient Estimates}

\begin{figure}[h!]
	\centering
		\caption{Coefficient Tuning Plots}
	\label{figure:main}
	\includegraphics[scale = 0.6]{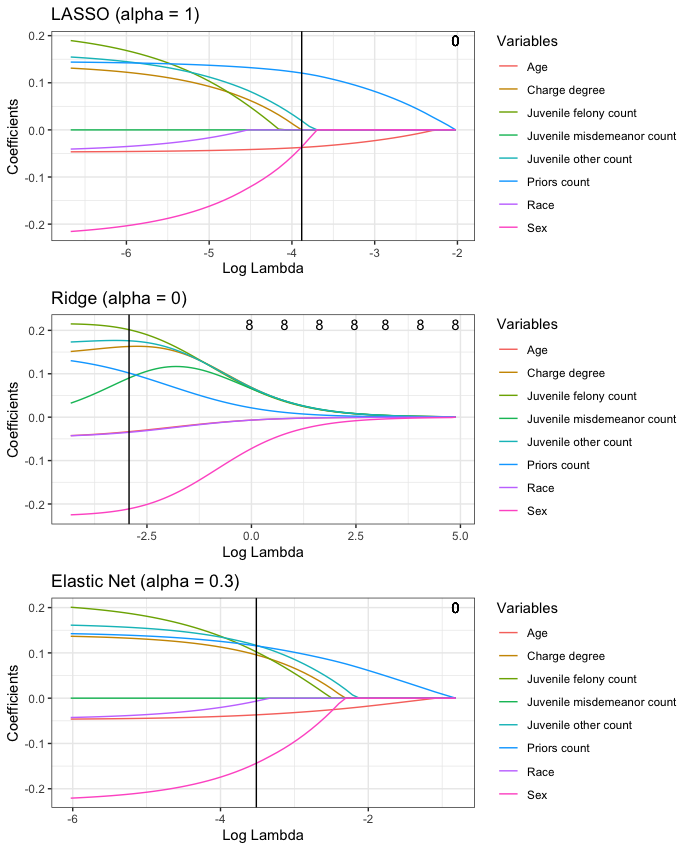}
\end{figure}

Next, we dig into the models to explore the tuned coefficient estimates across many model specifications. Here, we present individual fits of models but at various levels of penalty strength to explore the features that contributed to the most accurate specifications. The most intuitive way to do so is to visualize the tuning of the coefficient estimates across a range of logged $\lambda$ values. This allows for concise visual inspection, and thus interpretation of the mechanics of the penalization process at the feature level. The results are presented in Figure \ref{figure:main}, where solid vertical lines represent the optimal value of $\lambda$ used in the coefficient reports in Table \ref{table:coefs} in the Appendix. 

Recall that by penalizing regression coefficients that contribute to model instability (and thus inefficiency), we are able to hone in on the ``most important'' features. In this third stage, then, we address this issue across our three models, each with a different penalty strength. Overall we found that most of the features do appear to matter in predicting an individual's likelihood of recidivating, a finding divergent from Dressel and Farid (\citeyear{dressel2018accuracy}. Notably, Dressel and Farid found that only two features (\textit{age} and \textit{total prior convictions}) contribute to optimal classification solutions compared to the more complex COMPAS algorithm incorporating 137 features. However, we found that in the elastic net model, which was most accurate of our three models and more accurate than Dressel and Farid's linear \textit{and} nonlinear classifiers, the only dropped feature was \textit{juvenile misdemeanor counts}.\footnote{Importantly, Dressel and Farid's (2018) classifiers prevented the statistical dropping of features. As such, we make this point merely to demonstrate how our approach benefitted from and thus builds on Dressel and Farid's (2018) findings.} Our findings suggest that this feature was the only one that failed to contribute to unique explained variance. All other features in the elastic net model remained, suggesting that \textit{some} additional model complexity contributes to greater accuracy. 

The more extreme case of the LASSO, which penalizes the absolute value of the vector of coefficients seen in Equation \ref{eq:lasso}, dropped \textit{race}, \textit{juvenile felony count}, \textit{juvenile misdemeanor count}, and \textit{charge degree}, suggesting these features contribute to a decrease in model efficiency and stability. And as expected by Equation \ref{eq:ridge}, all features remained in the ridge case seen in the second plot in Figure \ref{figure:main}, though the tuning $\rightarrow$ 0 was closely mirrored by the other two penalized models.

\subsection{Checking Model Accuracy: ROC Curves}

\begin{figure}[h!]
	\centering
			\caption{ROC Curves for Each Model}
	\label{figure:roc}
	\includegraphics[scale = 0.55]{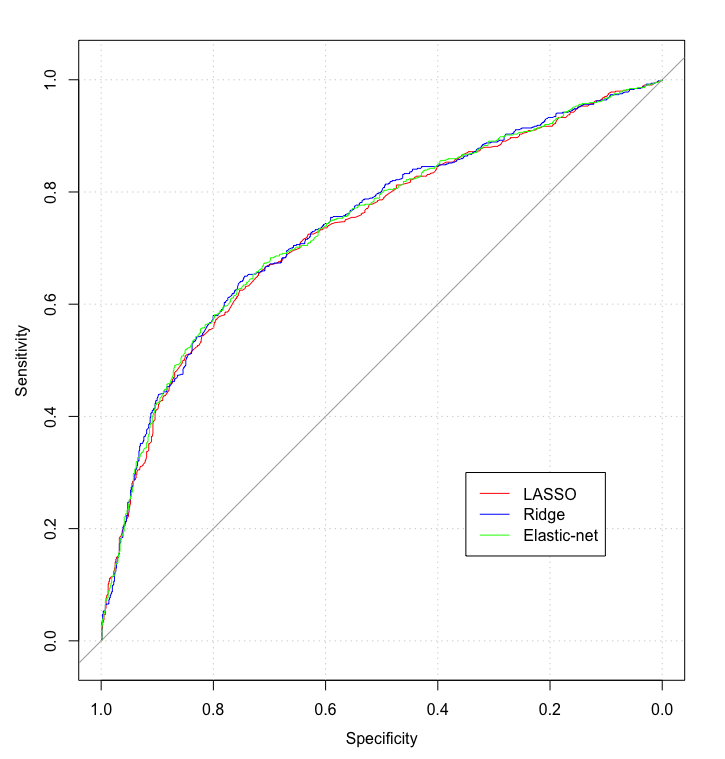}
\end{figure}

In the final stage, we move to a visual inspection of model-level accuracy using receiver operator characteristic (ROC) curves. The results are presented in Figure \ref{figure:roc}. For interpretation, curves pulled to the upper left of the plot suggest greater accuracy in classifying true positive rates relative to false positive rates. The farther the curve is to the upper left from the 45 degree line, the more accurate the results. Note each model has a unique line, denoted by color. 

Across all models, we see two things. First, the distance between the curves for all models and the 45 degree line is notable, pointing to relative accuracy. Second, and most importantly, there is strong agreement across all of our models, suggesting we are picking up true classification patterns in recidivism rates post-sentencing. To corroborate this visual pattern of consistency across the models, the areas under the curve (AUC) for the LASSO, ridge, and elastic net are 0.7323, 0.7388, and 0.7373 respectively.

\section{Discussion}

Building on the results in the previous subsections, all findings support our main premise that these relatively straightforward, accessible, open-source algorithms produced improved predictions over the proprietary, expensive alternative in COMPAS. In this section, we pull these findings apart to highlight their substantive contributions to our argument. 

In the first stage of the analysis we demonstrated that, as expected, the penalized regressions were about 2\% more accurate than the baseline COMPAS replication in predicting recidivism in the following two years for the Broward County cases. This finding is important because we were able to produce more accurate predictions with relative ease using computational tools available for public use. Further, recall the increase in statistical efficiency of the penalized regression approach, allowing for more appropriate selection of unique features. This point is addressed more below. Taken together, the first stage of the analysis strongly supports the value of open-source predictive algorithms in highly consequential, real-world contexts such as criminal sentencing. 

In the second stage, we showed the classification solutions, which emerged from our 1,000 model iterations in the first stage. Using confusion matrics, we showed that the overall classification rates were higher for true classes (negative and positive), compared to false classes (negative and positive). While our algorithms were by no means perfectly accurate in classification, building on the increase in overall predictive accuracy rates in the first stage, at a minimum our classification solution was more ``right'' than ``wrong''. The accuracy was corroborated using ROC curves and AUC calculations in the fourth section. These measures underscored the consistency. This relatively simple approach to predicting and classifying recidivism in a short period after sentencing suggests that researchers could significantly contribute to the development of algorithms, in that there is still much ground to be gained in strengthening the predictive accuracy of these models. An open-source framework would facilitate wider engagement to allow for such development.

In the third stage, we were interested in diving deeper into the predictive quality of the model to explore the features that contributed to the strongest predictive specifications. Recall that penalized regressions allow for a honing in on those features that contribute most uniquely to the best fitting model (i.e., the model with the least information loss). Thus, in the third stage, we built on the model-level predictive and classification results from the first two stages, and focused on feature-level nuances surrounding the ``best'' specifications. In so doing, we found that in the most accurate model (elastic net) all features except \textit{juvenile misdemeanor counts} contributed to increased accuracy in predicting recidivism at the optimal value of $\lambda$. 

Although the elastic net regression was most accurate, the LASSO and ridge models (displayed in the Appendix in Table \ref{table:coefs}) were within a range of 0.4\%. This is interesting, because the more extreme model (LASSO) dropped four features, and had an accuracy rate of only 0.2\% less than the elastic net. Though there is consistency across the models, the final group of features that contributed to the ``best'' model is less clear. In other words, the ``best'' model itself remains unclear from this analysis, as all models, each of which retained a different configuration of features, had similar accuracy rates. Indeed, within a narrow window of accuracy, there could realistically be a combination of features between four (LASSO) and eight (ridge) that contribute to some optimal model. As such, our results, in combination with Dressel and Farid's (2018) conclusion of only \textit{two} features mattering, point to a degree of variability in the precise set of features that contribute to the most accurate classifer. Such a finding underscores the need for investment in future and wider development of recidivism prediction algorithms, which is realistic and feasible in an open-source environment. 

In light of these results, we contend that even if our findings were identical with the COMPAS', judicial decision makers should nonetheless opt for the open-source, inexpensive choice for reasons of statistical efficiency, predictive accuracy, interpretability, and ultimately greater transparency. Our analysis is a conservative yet supportive starting place for our central argument, which is that better predictions with relatively minimal effort are possible. The implication, then, is scaling up the open source development of these algorithms could produce even better predictions of criminal recidivism. The end result would be more targeted judicial decision making, appropriate sentencing, and transparency in consequential contexts, all at relatively low costs to the users and decision makers. 

\section{Conclusion}

We have presented a series of theoretical and quantitative arguments that demonstrate open-source collaboration is preferable to closed-source development of algorithmic recidivism prediction tools. Building on the literature showing that proprietary algorithms may perpetuate harmful racial biases (Dressel and Farid 2018), transparent open-source development can help identify and fix these issues, in addition to contributing to better predictions and lowering costs of judicial decision making. 

In contexts as significant as setting bail or sentencing, which have serious effects on human lives, we ought to pool intellectual resources from methodological and domain-area experts to ensure the criminal justice system does not deploy computational tools that produce unjust outcomes, or proceed down inefficient, wasteful, and less accurate avenues. We have demonstrated that simple, straightforward open-source approaches to predictive risk assessment are much lower cost than expensive proprietary solutions, and can even produce more accurate predictions. We therefore strongly advocate for open-source algorithm development in consequential social contexts, especially in the realm of criminal sentencing.

\clearpage

\section{Appendix}

\subsection*{Coefficient Estimates Across Three Penalized Regressions}

Below we present the coefficient estimates from all three penalized regressions in the main paper, which are visually represented in Figure \ref{figure:main}. This is a valuable step in that, building on Dressel and Farid (\citeyear{dressel2018accuracy}), we offer a statistically efficient method for honing in on the most valuable features, resulting in well-fitting models that side step the threat of overfitting and complexity. 

We differ slightly from Dressel and Farid in that a few more features contributed to a well-fit model. Whereas the authors suggested only two features can produce as accurate predictions as the COMPAS algorithm, we found that seven features more contribute to better fitting model with the highest degree of accuracy seen in the elastic net specification in column three in Table \ref{table:coefs}. Upon fitting these models, they were then used to classify criminals who recidivated and those who did not, in line with the main goal of the paper. See the features that matter most in contributing to the best model fit, as well as their magnitudes of effect presented in Table \ref{table:coefs}. 

	\begin{table}[h!]
	\begin{center}
		\caption{Coefficient Results for Representative Regressions}
		\label{table:coefs}
		\begin{tabular}{ l c c c }
			\multicolumn{3}{r}{\textit{Model}} & \\
			\hline
			& & & \\
			& LASSO & Ridge & Elastic Net \\
			\hline
			\hline
			& & & \\
			\textit{Tuning parameters} & & & \\
			\cline{1-1}
			 & $\lambda_{1SE}$ = 0.021 & $\lambda_{1SE}$ = 0.053 & $\lambda_{1SE}$ = 0.030 \\
			 & $\alpha$ = 1 & $\alpha$ = 0 & $\alpha$ = 0.30 \\
			& & & \\
			\textit{Covariates} & & & \\
			\cline{1-1}
			Intercept & 0.682 & 0.574 & 0.638 \\
			Race & -- & -0.035 & -0.007 \\
			Sex & $-$0.036 & -0.211 & -0.144 \\
			Age & $-$0.037 & -0.033 & -0.037 \\
			Juvenile felony count & -- & 0.202 & 0.102 \\
			Juvenile misdemeanor count & -- & 0.090 & -- \\
			Juvenile other count & 0.019 & 0.176 & 0.117 \\
			Priors count & 0.121 & 0.102 & 0.116 \\
			Charge degree & -- & 0.163 & 0.096 \\
			& & & \\
			Averaged Test Accuracy Rate & 69.8\% & 69.6\% & 70\% \\
			\hline
			\multicolumn{4}{l}{\textit{Coefficients denoted by ``--" have been penalized to take on a value of 0, i.e., ``dropped''.}}
		\end{tabular}
	\end{center}
\end{table}

\clearpage

\subsection*{Replicating Dressel's Logistic Regression Classifier}

To further justify our approach of building on Dressel and Farid \citep{dressel2018accuracy}, we replicated Dressel's \citep{dressel2017accuracy} logistic regression classifier and found nearly identical results using the same features and data. We fit 1,000 iterations of the classifier and found a mean accuracy of 0.676 (67.6\%), while Dresel had a mean accuracy rate of 0.67 (67.0\%). This suggests that our justification for building on her approach using a penalized regression framework was not only statistically efficient (i.e., avoiding model complexity and overfitting), but was also reliable and robust to past findings.

\clearpage

\subsection*{Cross-Validation Results for Optimal Lambda Values}

\begin{figure}[h!]
	\centering
	\caption{Optimal Values of $\lambda$ Across All Models}
	\label{figure:lambda}
	\includegraphics[scale = 0.5]{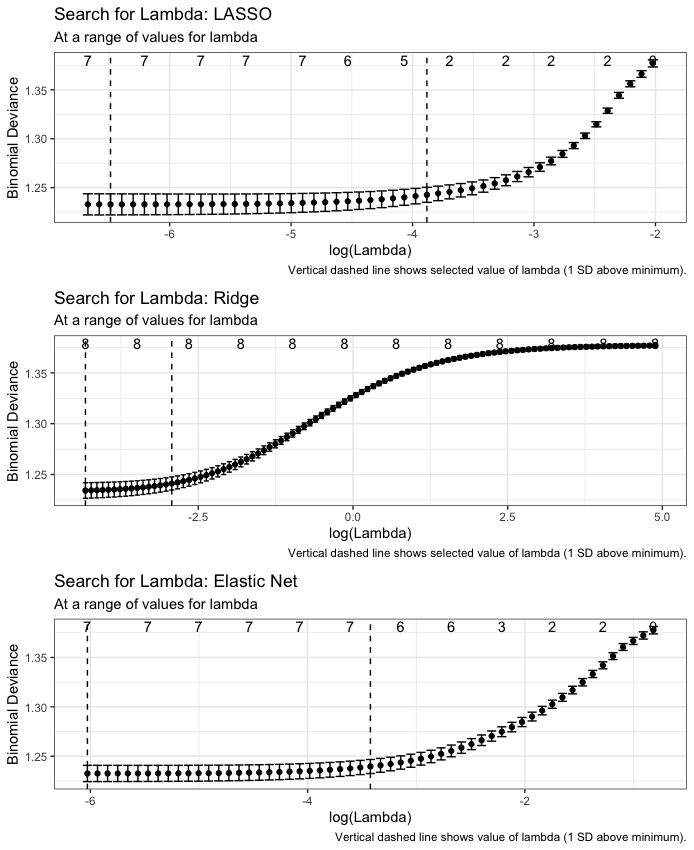}
\end{figure}

Recall that the $\lambda$ parameter controls the intensity of penalty imposed on the coefficient estimates. This is the ``tuning'' discussed throughout. The optimal method for selecting a good $\lambda$ value is that which minimizes information loss. Though some opt for $\lambda$ at the global minimal mean squared error (MSE), we opt for selecting $\lambda$ 1 standard deviation from $\lambda$ at the global minimum MSE, in line with popular wisdom \citep{hastie2014glmnet}. The results are presented in Figure \ref{figure:lambda}.

\clearpage

\subsection*{Mean Squared Error across Various Levels of $\alpha$}

Here we present the mean squared errors at multiple values of the mixing parameter $\alpha$ to demonstrate the reduction in loss by each specification of the model. For reference, when $\alpha = 0$, this is a ridge regression, whereas when $\alpha = 1$, the model is a LASSO regression. Recall that the $\alpha$ parameter controls the degree of blending these two extreme penalies. Thus, based on cross-validation tests, we found the optimal blending for the elastic net regression to be $\alpha = 0.3$, meaning the model should more closely resemble the ridge regression, compared to the LASSO regression. This is clearly seen in the table of results in the appendix in Table \ref{table:coefs} as well as in the figure of tuned coefficient estimates in Figure \ref{figure:main} in the main paper.  

\begin{table}[!htbp] \centering 
	\caption{Mean Squared Error across Various Levels of $\alpha$} 
	\label{table:mse} 
	\begin{tabular}{@{\extracolsep{5pt}} lc} 
		\\[-1.8ex]\hline 
		\hline \\[-1.8ex] 
		Alpha=0 (Ridge)& $0.2160484$ \\ 
		Alpha=0.1 & $0.2168683$ \\ 
		Alpha=0.2 & $0.2169188$ \\ 
		Alpha=0.3 (Elastic Net) & $0.2167233$ \\ 
		Alpha=0.4 & $0.2171698$ \\ 
		Alpha=0.5 & $0.2169559$ \\ 
		Alpha=0.6 & $0.2168203$ \\ 
		Alpha=0.7 & $0.2179104$ \\ 
		Alpha=0.8 & $0.2163541$ \\ 
		Alpha=0.9 & $0.2163106$ \\ 
		Alpha=1 (LASSO) & $0.2172665$ \\ 
		\hline 
		\hline \\[-1.8ex] 
	\end{tabular}
\end{table} 

\clearpage

\bibliographystyle{apsr}
\bibliography{algs}

@article{tibshirani1996regression,
	title={Regression shrinkage and selection via the lasso},
	author={Tibshirani, Robert},
	journal={Journal of the Royal Statistical Society: Series B (Methodological)},
	volume={58},
	number={1},
	pages={267--288},
	year={1996},
	publisher={Wiley Online Library}
}

@article{hastie2014glmnet,
  title={Glmnet vignette},
  author={Hastie, Trevor and Qian, Junyang},
  journal={Retrieve from http://www. web. stanford. edu/\~{} hastie/Papers/Glmnet\_Vignette. pdf.},
  volume={20},
  pages={2016},
  year={2014}
}

@article{dressel2018accuracy,
	title={The accuracy, fairness, and limits of predicting recidivism},
	author={Dressel, Julia and Farid, Hany},
	journal={Science advances},
	volume={4},
	number={1},
	pages={eaao5580},
	year={2018},
	publisher={American Association for the Advancement of Science}
}

@article{dressel2017accuracy,
	title={Accuracy and Racial Biases of Recidivism Prediction Instruments},
	author={Dressel, Julia J},
	year={2017}
}

@article{zou2005regularization,
	title={Regularization and variable selection via the elastic net},
	author={Zou, Hui and Hastie, Trevor},
	journal={Journal of the royal statistical society: series B (statistical methodology)},
	volume={67},
	number={2},
	pages={301--320},
	year={2005},
	publisher={Wiley Online Library}
}

@article{hoerl1970ridge,
	title={Ridge regression: Biased estimation for nonorthogonal problems},
	author={Hoerl, Arthur E and Kennard, Robert W},
	journal={Technometrics},
	volume={12},
	number={1},
	pages={55--67},
	year={1970},
	publisher={Taylor \& Francis Group}
}

@article{park2008bayesian,
	title={The bayesian lasso},
	author={Park, Trevor and Casella, George},
	journal={Journal of the American Statistical Association},
	volume={103},
	number={482},
	pages={681--686},
	year={2008},
	publisher={Taylor \& Francis}
}

@article{orsenigo2012kernel,
	title={Kernel ridge regression for out-of-sample mapping in supervised manifold learning},
	author={Orsenigo, Carlotta and Vercellis, Carlo},
	journal={Expert Systems with Applications},
	volume={39},
	number={9},
	pages={7757--7762},
	year={2012},
	publisher={Elsevier}
}

@inproceedings{wu2010heterogeneous,
	title={Heterogeneous feature selection by group lasso with logistic regression},
	author={Wu, Fei and Yuan, Ying and Zhuang, Yueting},
	booktitle={Proceedings of the 18th ACM international conference on Multimedia},
	pages={983--986},
	year={2010},
	organization={ACM}
}

@misc{angwin2016machine,
	title={Machine Bias: there’s software used across the country to predict future criminals. And it’s biased against blacks. ProPublica 2016},
	author={Angwin, Julia and Larson, Jeff and Mattu, Surya and Kirchner, Lauren},
	year={2016}
}

@article{kennedy2018trust,
	title={Trust in Public Policy Algorithms},
	author={Kennedy, Ryan and Waggoner, Philip and Ward, Matthew},
	journal={Available at SSRN 3339475},
	year={2018}
}

@inproceedings{green2019disparate,
	title={Disparate interactions: An algorithm-in-the-loop analysis of fairness in risk assessments},
	author={Green, Ben and Chen, Yiling},
	booktitle={Proceedings of the Conference on Fairness, Accountability, and Transparency},
	pages={90--99},
	year={2019},
	organization={ACM}
}

@article{eckhouse2019layers,
	title={Layers of bias: A unified approach for understanding problems with risk assessment},
	author={Eckhouse, Laurel and Lum, Kristian and Conti-Cook, Cynthia and Ciccolini, Julie},
	journal={Criminal Justice and Behavior},
	volume={46},
	number={2},
	pages={185--209},
	year={2019},
	publisher={SAGE Publications Sage CA: Los Angeles, CA}
}

@article{olhede2018growing,
	title={The growing ubiquity of algorithms in society: implications, impacts and innovations},
	author={Olhede, SC and Wolfe, PJ},
	journal={Philosophical Transactions of the Royal Society A: Mathematical, Physical and Engineering Sciences},
	volume={376},
	number={2128},
	pages={20170364},
	year={2018},
	publisher={The Royal Society Publishing}
}

@article{holsinger2018rejoinder,
	title={A Rejoinder to Dressel and Farid: New Study Finds Computer Algorithm Is More Accurate than Humans at Predicting Arrest and as Good as a Group of 20 Lay Experts},
	author={Holsinger, Alexander M and Lowenkamp, Christopher T and Latessa, Edward and Serin, Ralph and Cohen, Thomas H and Robinson, Charles R and Flores, Anthony W and VanBenschoten, Scott W},
	journal={Fed. Probation},
	volume={82},
	pages={50},
	year={2018},
	publisher={HeinOnline}
}

@article{waggonerbig,
	title={Big Data and Trust in Public Policy Automation},
	author={Waggoner, Philip D and Kennedy, Ryan and Le, Hayden and Shiran, Myriam},
	journal={Statistics, Politics and Policy},
	publisher={De Gruyter}
}

\end{document}